\documentclass[twocolumn,conference]{IEEEtran}
\usepackage[T1]{fontenc}
\usepackage[latin9]{inputenc}
\usepackage{amsmath}
\usepackage{graphicx}
\usepackage[unicode=true,
 bookmarks=true,bookmarksnumbered=true,bookmarksopen=true,bookmarksopenlevel=1,
 breaklinks=false,pdfborder={0 0 1},backref=false,colorlinks=false]
 {hyperref}
\hypersetup{pdftitle={Your Title},
 pdfauthor={Your Name},
pdfborderstyle=,pdfpagelayout=OneColumn,pdfnewwindow=true,pdfstartview=XYZ,plainpages=false}

\makeatletter

\providecommand{\tabularnewline}{\\}

\usepackage{array}
\providecommand{\tabularnewline}{\\}

\usepackage{paralist}

\usepackage{relsize}
\usepackage{xcolor}
\usepackage{cite}
\usepackage[normalem]{ulem}
\usepackage{url}
\usepackage{tikz}
\usepackage[font=footnotesize]{caption}
\usepackage{subcaption}

\newcommand\copyrighttext{Published at the 8th IEEE International Workshop on Sensors and Smart Cities (SSC 2022), hosted by the 2022 IEEE International Conference on Smart Computing (SMARTCOMP 2022).}
\newcommand\copyrightnotice{%
\begin{tikzpicture}[remember picture,overlay]
\node[anchor=south,yshift=30pt] at (current page.south) {\fbox{\parbox{\dimexpr\textwidth-\fboxsep-\fboxrule\relax}{\copyrighttext}}};
\end{tikzpicture}%
}

\makeatother

\begin{document}
\IEEEoverridecommandlockouts
\title{Performance Evaluation of Switching Between WiFi and LiFi under a
Common Virtual Network Interface\vspace{-2mm}}
\author{\IEEEauthorblockN{Loreto~Pescosolido, Emilio Ancillotti, Andrea Passarella}\IEEEauthorblockA{Italian National Research Council, Institute for Informatics and Telematics
(CNR-IIT)\\
 Via Giuseppe Moruzzi 1, 56124 Pisa, Italy\\
 Email: \{loreto.pescosolido, emilio.ancillotti, andrea.passarella\}@iit.cnr.it\vspace{-6mm}
 }}

\maketitle
\copyrightnotice
\begin{abstract}
We consider a hybrid wireless local area network composed of both
WiFi and LiFi Access Points (AP) and wireless devices. Each device
is identified in the network by a unique IP address, using a virtual
network interface obtained by bonding the WiFi and LiFi physical interfaces,
implemented through commercially available products. We measure the
time it takes to switch between the two \emph{physical} interfaces
and its impact on the traffic flow, under different settings of the
mechanisms used by the interface bonding driver. Different specific
triggering events are considered for the switch, namely: an (simulated)
interface malfunctioning or unintended shutdown, a signal loss, and
a manual (intended) switch. Our experimental results show that the
different types of triggering events have an impact on the time it
takes to reconfigure the currently active physical interface (which
is used by the virtual interface to send/receive data), with connection
recovery times ranging from few tens milliseconds to few seconds.
This entails a packet loss on active flows which, in the worst case,
we quantify in a maximum loss of up to 1\% of the traffic flowing
during 1 second. 
\end{abstract}

%\vspace{3mm}


%\IEEEpubid{\makebox[\columnwidth]{XXXXXXXXXXXXXXXX/\$31.00 \copyright 2013 IEEE \hfill} \hspace{\columnsep}\makebox[\columnwidth]{ }}

\section{Introduction}

Energetically sustainable smart home and smart building appliances,
ambient assisted living, health and lifestyle monitoring and assistance,
and home automation appliances are expected to contribute to an increase
in the number and variety of IoT devices in use in indoor wireless
local area networks. IoT devices can be used to interact with the
user and collect data for applications pertaining the users\textquoteright{}
own scope, but they can also provide meaningful information to the
smart city environment. For instance, electricity and water consumption
monitoring tools are a useful source of data in smart buildings for
sustainable smart cities. Monitoring the users\textquoteright{} behavior
indoor (home, office, restaurants, etc..) can help predict, for instance,
user movements across different areas of the city, and so on. Increasing
the connectivity and available bandwidth in indoor environments is
therefore of paramount importance to support this kind of ever increasing
traffic.

In this context, the outstanding results achieved in the recent years
by the research in the area of Light-Fidelity (LiFi) \cite{Haas2016}
suggest that infrared/visible light communications can be a viable
tool to increase the overall avaliable bandwidth and reliability of
indoor communications for IoT devices. Commercial LiFi transducers,
and LiFi-based LANs, have become available as well in the recent years,
although not yet at prices comparable to their traditional WiFi counterparts
and not yet able to exploit, in terms of data rate, the enormous potential
bandwidth offered in the VLC and infrared spectrum. This includes
both Access Points (APs) using led lamps as transmitting transducers
and infrared (IR) sensors as receiving ones, as well as USB network
interfaces using IR LEDs as transmitters and and photo-diodes as receivers.
Although able to provide a self contained LAN environment, it has
become clear that a more effective way of using the newly available
technologies is to do this in conjunction with communication systems
operating in different bands. In fact, the coverage radius of a typical
attocell (the area covered by a single access point) is in the order
of few (in the order of 5) meters, which (among other reasons), depending
on the scenario, may call for the presence of multiple parallel technologies
to form the LAN. Thus, hybrid LiFi/WiFi networks \cite{Zeng2020}
(see also the references therein) have emerged as a promising wireless
networking paradigm for many scenarios. 

The integration of wireless communication and networking capabilities
of LiFi and WiFi may be performed at different layers of the communication
protocol stack. At layers above the TCP/IP layer, it entails that
each physical network interface on the same device is mapped to a
different IP address. Possibly, it may also belong to a different
subnet. This requires that the handling of the multiple interfaces,
i.e., selecting which interface to use to forward outgoing traffic,
is performed at layers above the TCP/IP. This type of configuration
has been used in several works, e.g. \cite{Zeng2020}, for the purpose
of studying the overall system capacity, but it can be argued that
it is not an ideal choice in terms of complexity added to the upper
layers, which should track the currently active IP address of any
device, among the multiple ones each device can use. In fact, integration
at the lower layers of the protocol stack has been recently receiving
attention. In \cite{Zubow2021}, the integration is performed at the
physical layer in the form of extending 802.11 COTS by attaching a
VLC transducer to one of the antenna ports, in order to obtain a completely
transparent network interface. Although proving a promising solution,
this approach has not yet led to commercially available products.

In this work, we consider the integration of a LiFi and a WiFi network
interfaces at layer 2 (data link), building a single ``virtual''
network interface which combines WiFi and LiFi COTS without any hardware
modification. To do this we exploit the Linux Ethernet Bonding Driver
\cite{linuxbonding}, a module of the Linux Kernel originally designed
for bonding multiple ethernet interfaces which, however, can even
be used with wireless interfaces.

We present a set of experimental results aimed at quantifying the
connectivity downtime when there is a switch between the LiFi and
WiFi physical interface as the consequence of an event that can be
either exogenous (signal or carrier loss, physical interface failure,
etc...) or intentional, e.g., a switch operated for load balancing
purposes. Based on the results, we discuss the pros and cons of several
settings of the bonding driver. Particularly, we focus on the means
and sampling frequency used by the bonding driver to check the status
of the physical links corresponding to the two interfaces.

The rest of the paper is organized as follows, in Section~\ref{sec:Virtual-common-interface}
we describe the functions of the software tool we used to combine
the WiFi and LiFi interface into a single virtual network interface
for the wireless devices, and the related configuration of the wireless
LAN. In Section~\ref{sec:Testbed description}, we describe the testbed
we used to perform our experiments, and present the methodology to
extract the relevant information about the interface switching time
from the experimental traces. In Section~\ref{sec:experimental-results}
we present our experimental results and discuss them with the aim
of obtaining indications on how hybrid networks can impact on the
QoS of different types of traffic in different scenarios. Finally,
Section~\ref{sec:Conclusion} concludes the paper, summarizing our
contribution and take home message.\texttt{\vspace{-1mm}
}

\section{Virtual common interface\label{sec:Virtual-common-interface}\texttt{\vspace{-2mm}
}}

We refer to a virtual common interface as an entity that appears to
the operating system (OS) of a wireless device as a fully effective
network interface, to which it is assigned a single IP address, but,
at layers below the TCP/IP, it relies on multiple (in our case, two)
physical network interfaces. To implement the virtual common wireless
network interface we exploit the Linux Ethernet Bonding Driver \cite{linuxbonding}.
The main features of the bonding driver are summarized below, along
with the description of how it can be used to handle a WiFi and a
LiFi interface in a wireless LAN, and some necessary additional detail
on the mechanisms the driver uses to check the status of the two interfaces,
in order to react to changing conditions.\texttt{\vspace{-1mm}
}

\subsection{The Linux Ethernet Bonding Driver\texttt{\vspace{-1mm}
}}

The driver has been part of the Linux Kernel from its early stage
of development. The last version of the driver is 2.6 \cite{linuxbonding}.
The driver was designed to handle multiple Ethernet network interfaces
under a virtual network interface. This allows to present the device
to the network under a unique IP address, thus making it transparent,
to the TCP/IP and upper layers of both the considered device and the
other network devices, the presence of different means for letting
traffic reach, or depart from, the considered device. Although the
driver was designed to handle Ethernet interfaces, some of its functions
can be used with wireless interfaces as well.

The driver allows to chose among several policies, called ``modes'',
for distributing traffic across the available interfaces. Some of
these policies are oriented to load-balancing, other policies are
more oriented to increase the system reliability. However, the possibility
to enable a given policy is based on some requirements on the network
interfaces in use. Particularly, many policies require that the physical
interfaces support interface the ``ethtools'' library. While this
support is a standard feature of any Ethernet interface, it is typically
absent from wireless interfaces. This limits the range of policies
that can be implemented by the driver when, as in the case of the
LiFi and WiFi interfaces considered in this work, the ``ethtools''
support is not available. Fortunately, for the purposes of this work,
the use of modes not requiring the ``ethtools'' support is sufficient.
Particularly, in the driver configuration, we have considered the
\texttt{active-backup} mode selection. In the \texttt{active-backup}
mode, one of the interfaces is declared as the \emph{default} one,
and it is used by the device for outgoing or incoming traffic whenever
available . However, when the interface or its corresponding physical
link is not available, the driver switches to the other physical interface,
called the \emph{backup} one \cite{linuxbonding}.

The presence and correct operation, in an IP subnet, of devices utilizing
bond interfaces, relies on a correct utilization of the address resolution
protocol (ARP) \cite{ARP-cisco}, as it is crucial to keep track of
the association, at any given time, between IP addresses and MAC addresses
of the physical interfaces. With bond interfaces, this association
may vary much more frequently than in traditional networks where IP
and MAC addresses are mapped one-to-one in a static way. Moreover
(see below) ARP messages \emph{can} be also used by the bonding driver
as a means to proactively update the primary interface upon detection
of a link or interface failure in the \texttt{active-backup} mode,
even in the absence of traffic.\texttt{\vspace{-1mm}
}

\subsection{ARP and MII monitoring of the physical link\texttt{\vspace{-1mm}
}}

In the \texttt{active-backup} mode of the bonding driver, to keep
track of, and if necessary, switch the status of the interface, the
bonding driver performs periodical checks using either of two mechanisms,
called ARP monitoring and MII monitoring.

ARP monitoring uses standard ARP messages. A device in which bonding
is in operation, periodically broadcasts ARP request packets using
the currently active physical interface. The interface broadcasts
it through the network indicating a queried IP address (called \texttt{arp\_ip\_target}),
or even more than one. Typically, for the purpose of ARP monitoring
operation, this is the IP address of a designed device in the network.
The bond interface driver waits for a suitable time interval to receive
ARP reply packets, whose reception confirms to the driver that the
physical interface is working properly. A driver parameter that may
have an impact on the performance is the \texttt{arp\_validate} parameter.
This is used to select what types of ARP (or even non-ARP) packets
are used to determine the status of an interface when ARP monitoring
is used (more details in \cite{linuxbonding}). In the experiments
described in this work, \texttt{arp\_validate} was set to the value
3, which means that an interface is considered to be active only based
on ARP replies from the \texttt{arp\_ip\_target}. In practice, if,
after a suitable timeout, no ARP replies from the devices designed
by the \texttt{arp\_ip\_target} IP address are received, the interface
(and the link) is considered to be down, and a switch to the backup
interface is performed. An examination of the effect of different
settings of this parameter is outside the scope of this work, while
it will be considered in our future works.

With MII monitoring, no messages are sent through the network for
interface status monitoring purposes. Instead, to check he link status,
the driver only queries, internally, the currently active physical
interface. The physical interface own driver is in charge of monitoring
the physical availability of the (wired or wireless) link. An important
technical aspect, which has an impact on how the bonding driver works,
is how each physical interface keeps track of the status of its own
physical link. One desirable feature for the interfaces is that they
keep track of the status continuously, or periodically, sensing some
carrier signal, and that this signal can be queried by an external
driver using the \texttt{netif\_carrier\_ok()} function, which the
interface is required to support. Both the interfaces considered in
this work provide support to the \texttt{netif\_carrier\_ok()} function.
More details on other means for querying the status of the digital
link may be found in \cite{linuxbonding}.

Both the two types of monitoring (ARP and MII) present advantages
and disadvantages, in terms of overhead, reliability, and delay with
which a change in the link status is reflected in the physical interface
selection by the driver. Investigating these advantages and disadvantages
in a qualitative and quantitative way is the goal of the experimental
results we present in this work.

\texttt{\vspace{-8mm}
 }

\section{Testbed description\label{sec:Testbed description}\texttt{\vspace{-2mm}
}}

The testbed we used for this work consists of (i) a LiFi AP mounted
on the ceiling of an office room, at a 4m height, (ii) a wireless
device, namely, a PC-Stick, equipped with a WiFi internal interface
and an external USB LiFi interface, placed on a table, at 1m height,
under the coverage of the LiFi AP, (iii) a WiFi AP, placed on the
same table. A laptop PC, used to launch and control the experiments,
and a virtual machine running a DHCP server complete the set of devices.
The virtual machine runs a customized Linux family OS, provided by
the producers of the LiFi equipment. The LiFi and WiFi APs, and the
PC (and the DHCP virtual machine), were connected to an ethernet switch.
The LiFi and WiFi APs were configured as IP bridges, extending the
IP address space of a unique subnet to the wireless domain. Note that
the wireless device were mapped to single IP address, regardless of
the physical interface in use at any given time. The presence of a
dedicated virtual machine for the DHCP server was required to ease
the DHCP configuration by the tool provided with the LiFi components
by the producer.\texttt{\vspace{-2mm}
}

\subsection{Hardware\texttt{\vspace{-1mm}
}}

As wireless devices, we have used a PC-stick ADJ 270-00108 equipped
with an Intel Atom Z8350 processor, 2 GB RAM and 32 GB eMMC hard disk,
802.11 a/b/g/n/ac WiFi card, Bluetooth 4.0, 1 USB 2.0 port, 1 USB
3.0 port, 1 HDMI port. In the PC-Stick and the laptop PC, the Ubuntu
20.04 operating system is installed. The WiFi AP is a Gatework GW5300-Ventana
with an OpenWRT OS.

Both the LiFi APs (PureLifi LiFi-XC AP) LiFi USB connectors (IR/VLC
PureLifi LiFi-XC Station Dongle) are products of PureLifi \cite{PureLifi}.
The lamp is a 20W 4000 K Lucicup II by Lucibel, with a maximum luminous
power of 1930 lm.\texttt{\vspace{-2mm}
}

\subsection{Network configuration and management\texttt{\vspace{-1mm}
}}

Using the Linux Kernel bonding function we handle a LiFi and a WiFi
network interface under the same IP. In our setup, the two interfaces
used different MAC addresses (although the bonding function also allows
to handle interfaces with the same MAC address). Handling the different
MAC addresses at the network layer, i.e., insuring that at any given
time the traffic directed to device with a given IP address is forwarded
through the correct network path, is performed using the ARP signalling.

In bonding version 2.6.2 or later, when a failover or a change in
the currently active slave occurs in the active-backup mode, one or
more gratuitous ARP messages, determined by the \texttt{num\_grat\_arp}
parameter, are issued on the newly active slave interface. In our
testbed we have set the \texttt{num\_grat\_arp} parameter to 2 to
generate two gratuitous ARPs when the active slave change event occurs.

To measure the time it takes between events occurring on different
devices (interfaces or transducers shutdowns, active interfaces switches,
AP association, etc...) we set up an NTP server\footnote{The Network Time Protocol (NTP) is an Internet protocol for synchronizing
the clocks of hosts on the Internet with a granularity of few milliseconds.}, connected it to the Internet, and used it to keep the clocks of
the devices used in the experiments synchronized, by executing the
\texttt{ntpdate} command on every host at the beginning of each experiment
replica.

\vspace{-1mm}

\section{Experimental results\label{sec:experimental-results}\texttt{\vspace{-1mm}
}}

We conducted a set of experiments to test the performance of the bond
interface in the \texttt{active-backup} mode in terms of delay with
which the virtual interface switches from the primary interface to
the backup one. The experiments aim at determining the following quantities: 
\begin{description}
\item [{i)}] The time it takes to switch from the primary interface to
the backup one when the primary interface becomes unavailable, e.g.
due to a software or hardware problem. 
\item [{ii)}] The time it takes to switch from the primary interface to
the backup one when physical connectivity on the primary interface
drops due to a signal or carrier loss.
\item [{iii)}] The impact on a packet flow, in terms of packet loss percentage
(PLR), of an intentional interface switch. 
\end{description}
In all the experiments we evaluate the system performance by selecting
either LiFi or WiFi as the primary interface. In the experiments targeting
the switching delay, we consider both ARP and MII monitoring as the
monitoring tool. In the experiment targeting the PLR, we evaluate
both the downlink and uplink. 

Tables I and II show the bond interface configuration parameters in
use when ARP monitoring (Table~I) and MII monitoring (Table~II)
were used. The reader can refer to \cite{linuxbonding} for a more
detailed description of the meaning of each parameter.

Figures 1 through 4 show the histogram of the reaction time following
any event (interface shutdown, or carrier loss) after which the primary
interface is replaced by the backup one.

In all the experiments, a number of 600 event replicas were produced.
The histogram are normalized so that they can be interpreted as a
sample discrete probability density functions, i.e., the sum of the
areas of each bar equals one.%\texttt{\vspace{-1mm}}

\subsection{Switching delay after an interface shutdown%\texttt{\vspace{-1mm}}
}

The first set of results targets the time it takes to the interface
to switch from the active interface to the backup one, when the internal
communication between the device CPU and the physical network interface
of the device becomes unavailable, i.e, it becomes no more possible
to communicate with the interface on the device bus. This could be
the result, for instance, of a hardware or software failure. In our
experiment, we simulated such an event by sending a shut down command
to the said network interface, and measuring the time it takes for
the system to detect the unavailability and switch to the backup interface.
The results are collected in Figures~1 and 2.
\begin{table}[t]
\caption{Configuration parameters when using ARP monitoring}

\centering{}%
\begin{tabular}{|c|c|}
\hline 
parameter & value\tabularnewline
\hline 
\hline 
mode & \texttt{active-backup}\tabularnewline
\hline 
primary interface & LiFi or WiFi interface\tabularnewline
\hline 
\texttt{arp\_interval} & different values\tabularnewline
\hline 
\texttt{arp\_ip\_target} & IP target of the control PC\tabularnewline
\hline 
\texttt{arp\_validate} & 3\tabularnewline
\hline 
\texttt{fail\_over\_mac} & 1\tabularnewline
\hline 
\texttt{num\_grat\_arp} & 2\tabularnewline
\hline 
\end{tabular}\label{tab:ARP}
\end{table}
\begin{table}[t]
\begin{centering}
\caption{Configuration parameters when using MII monitoring}
\par\end{centering}
\centering{}%
\begin{tabular}{|c|c|}
\hline 
parameter & value\tabularnewline
\hline 
\hline 
mode & \texttt{active-backup}\tabularnewline
\hline 
primary interface & LiFi or WiFi interface\tabularnewline
\hline 
\texttt{miimon} & different values\tabularnewline
\hline 
\texttt{downdelay} & 0\tabularnewline
\hline 
\texttt{updelay} & 0\tabularnewline
\hline 
\texttt{fail\_over\_mac} & 1\tabularnewline
\hline 
\texttt{num\_grat\_arp} & 2\tabularnewline
\hline 
\end{tabular}\label{tab:MII}\vspace{-4mm}
\end{table}
\texttt{\vspace{-3mm}
}

\subsubsection{Switching delay after an interface shutdown with ARP monitoring}

\begin{figure}[t]
\begin{subfigure}{0.5\columnwidth}\includegraphics[width=0.95\columnwidth]{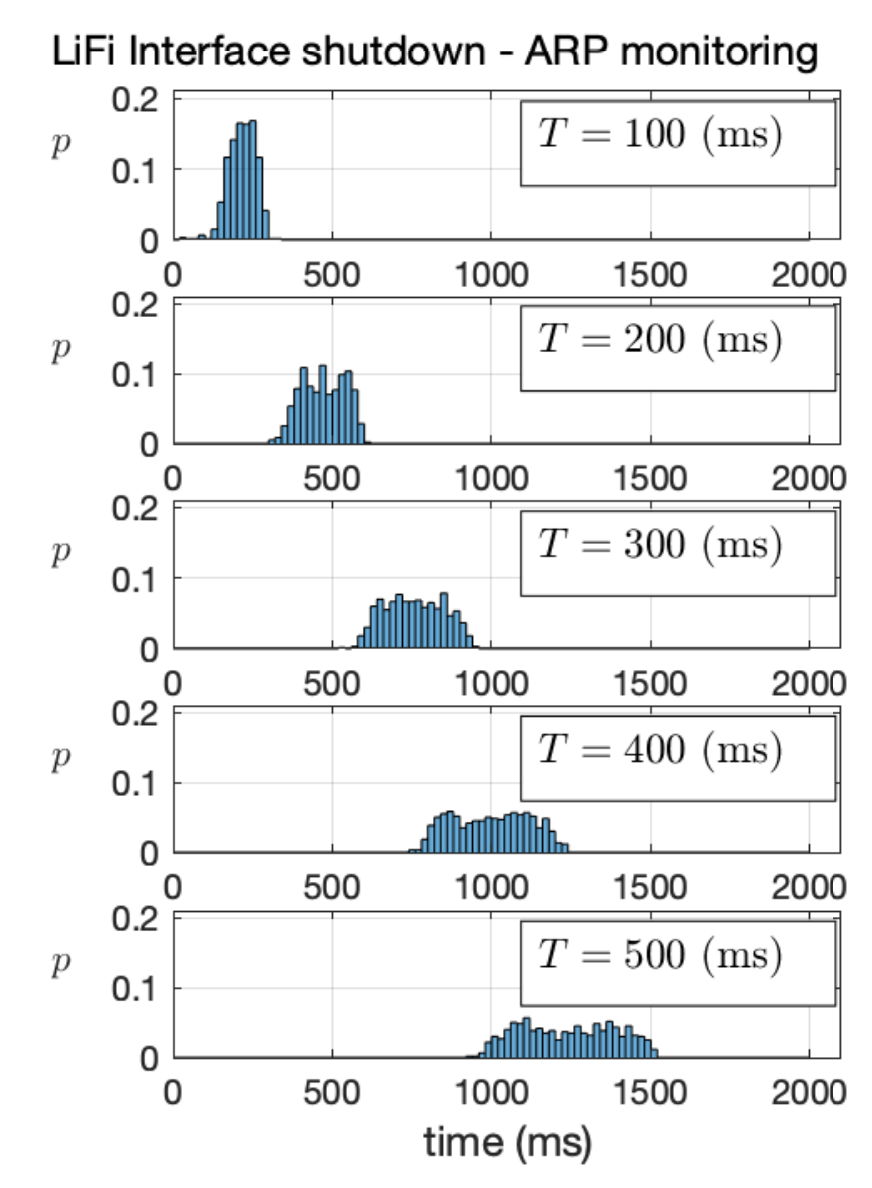}\caption{LiFi}\label{fig:1}\end{subfigure}\begin{subfigure}{0.5\columnwidth}\includegraphics[width=0.95\columnwidth]{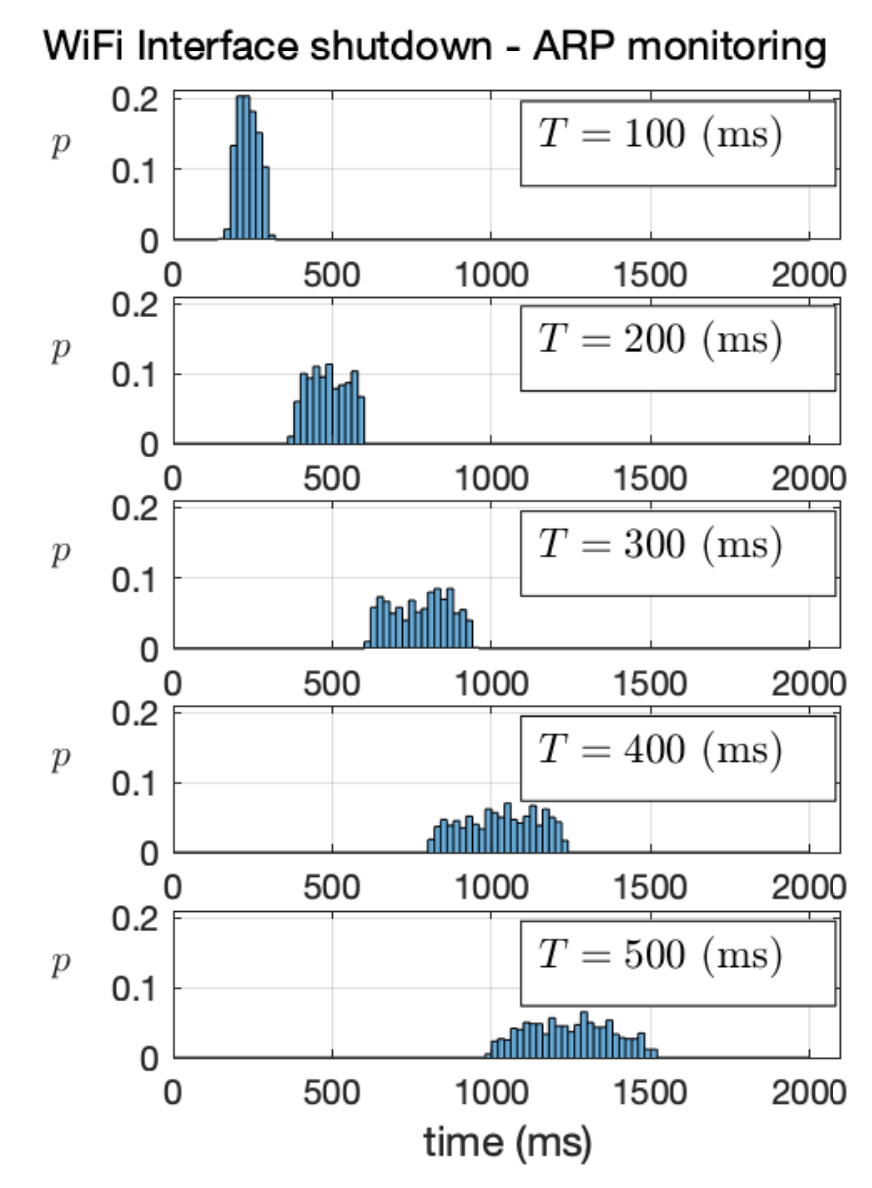}\caption{WiFi}\label{fig:2}\end{subfigure}

\caption{Reaction to a network interface shutdown with ARP monitoring}
\vspace{-3mm}
\end{figure}

In Subfigure~\ref{fig:1}, it can be seen that when the LiFi is set
as the active interface, switching to WiFi as the backup, as a consequence
of the LiFi interface becoming unavailable, requires a time between
200 and 300~ms when the ARP interval $T$ is set to 100~ms. The
range linearly increases up to 1--1.5 seconds when $T$ is set to
500~ms. The probability distribution of the delay is relatively uniform
between the interval edges, as the ARP sampling events and the interface
shutdown are not correlated.

In Figure~\ref{fig:2} we can see that the results obtained when
the active interface is the WiFi, and the device switches to LiFi
upon a WiFi interface shutdown, are quite similar. The reason why
the lower edge of the intervals increases with increasing $T$ is
that the system has to wait for a number of missed ARP replies before
declaring the currently active interface as unavailable, and this
time obviously scales up with $T$.

\subsubsection{Switching delay after an interface shutdown with MII monitoring}

\begin{figure}[t]
\begin{subfigure}{0.5\columnwidth}\includegraphics[width=0.95\columnwidth]{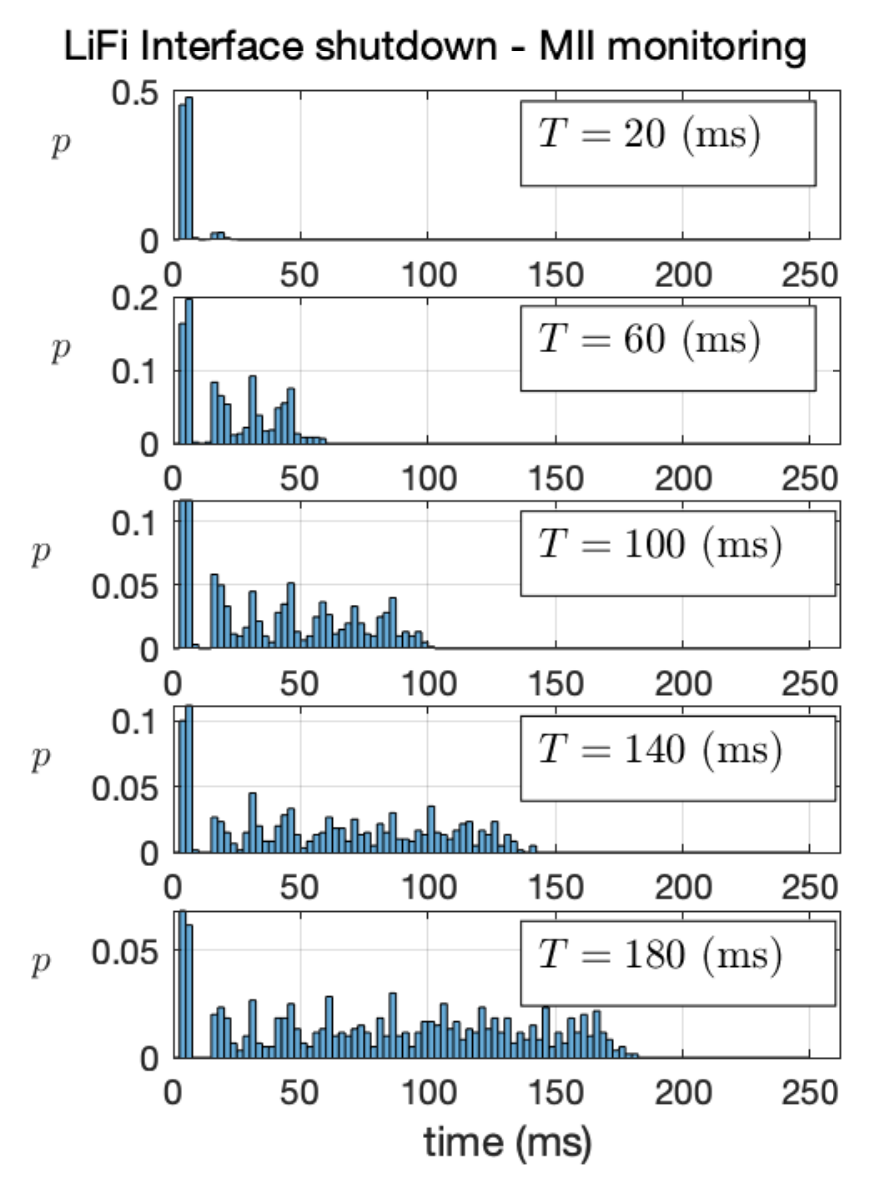}\caption{LiFi}\label{fig:3}\end{subfigure}\begin{subfigure}{0.5\columnwidth}\includegraphics[width=0.95\columnwidth]{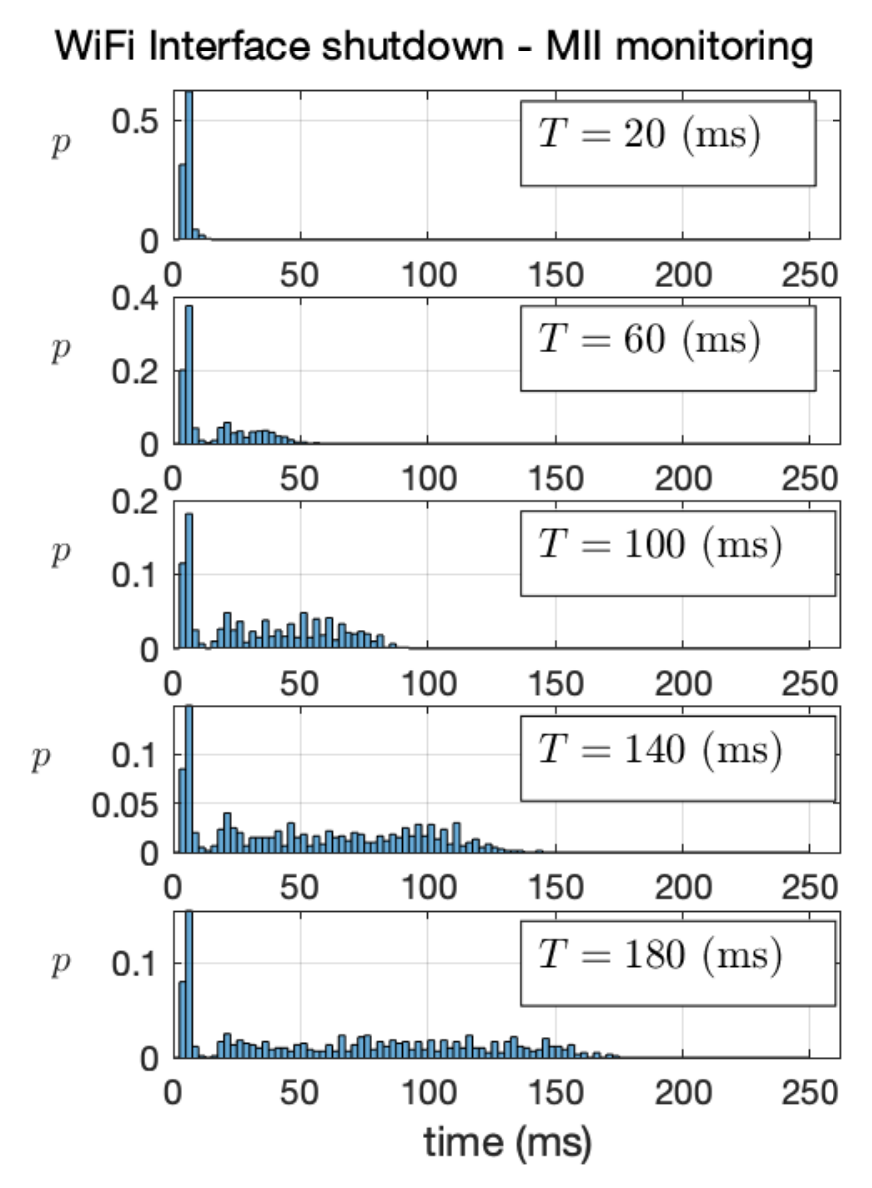}\caption{WiFi}\label{fig:4}\end{subfigure}

\caption{Reaction to a network interface shutdown with MII monitoring}
\vspace{-6mm}
\end{figure}

With MII monitoring, the time it takes to detect the shutdown and
switch to the backup interface is extremely lower, as showed in Subfigures~\ref{fig:3}
and~\ref{fig:4}. In fact, detecting a LiFi interface failure and
switching to the WiFi interface takes an interval in the range 5--20ms
with a MII monitoring interval $T$ of 20~ms, and 5--180~ms with
a 180ms interval. 
The results for the reverse switch, i.e., in response to a WiFi interface
shutdown, are quite similar. It can be noticed in Subfigure~\ref{fig:3}
that the interval samples are centered around some peaks. This peculiar
behavior can be likely attributed to the fact that the links status
in the LiFi interface might not be tracked continuously by its own
driver. Instead, it could be polled at regular intervals. This type
of update of the link status in the interface is indeed present in
some network interfaces, and was already taken into account at the
time the Linux bonding driver was developed, see \cite[Section 8.3]{linuxbonding}.
The distributions seen in Subfigure~\ref{fig:4}, referring to the
case when the system starts in the WiFi mode and switches to LiFi,
are more regular. However, from a macroscopic point of view, this
difference has basically no impact on the key finding, i.e., that
the delay with which the bond mechanism switches to the backup interface
is more or less uniformly distributed between a value close to zero
and the duration of the MII monitoring polling interval.

Indeed, the lower edge more or less coincides with the time to physically
execute the switch upon the detection of the negative \texttt{netif\_carrier\_ok()}
response. Note that, in this case, differently from the case of ARP
monitoring, the lower edge does not increase with $T$, as there is
no round-trip delay to wait in response to any signalling packet.\vspace{-2.5mm}

\subsection{Switching delay after a signal loss\vspace{-1.5mm}
}

In a real-life scenario, signal losses may be due to mobility, with
a device moving away from the coverage area of an AP, to air link
obstruction, or to interference. In our experiments however, to mimic
the effect of an abrupt signal loss for any of the two interfaces
on the wireless device, in order to fulfill the need for ms-level
precision and accuracy of the measurement of the switching delay (between
the triggering event and the completion of the switching operation),
and to perform a large number of replicas, we issued commands to either
turn down the lamp (when the active interface is the LiFi) or set
the power on the WiFi antenna to zero (when the active interface is
the WiFi). In Subsections~\ref{subsec:Switching-delay-after-signal-loss-ARP}
and \ref{subsec:Switching-delay-after-signal-loss-MII} we present
the measurements results. The results are collected in Figures~3
and 4.

\subsubsection{Switching delay after a signal loss - ARP monitoring\label{subsec:Switching-delay-after-signal-loss-ARP}}

\begin{figure}[t]
\begin{subfigure}{0.5\columnwidth}\includegraphics[width=0.95\columnwidth]{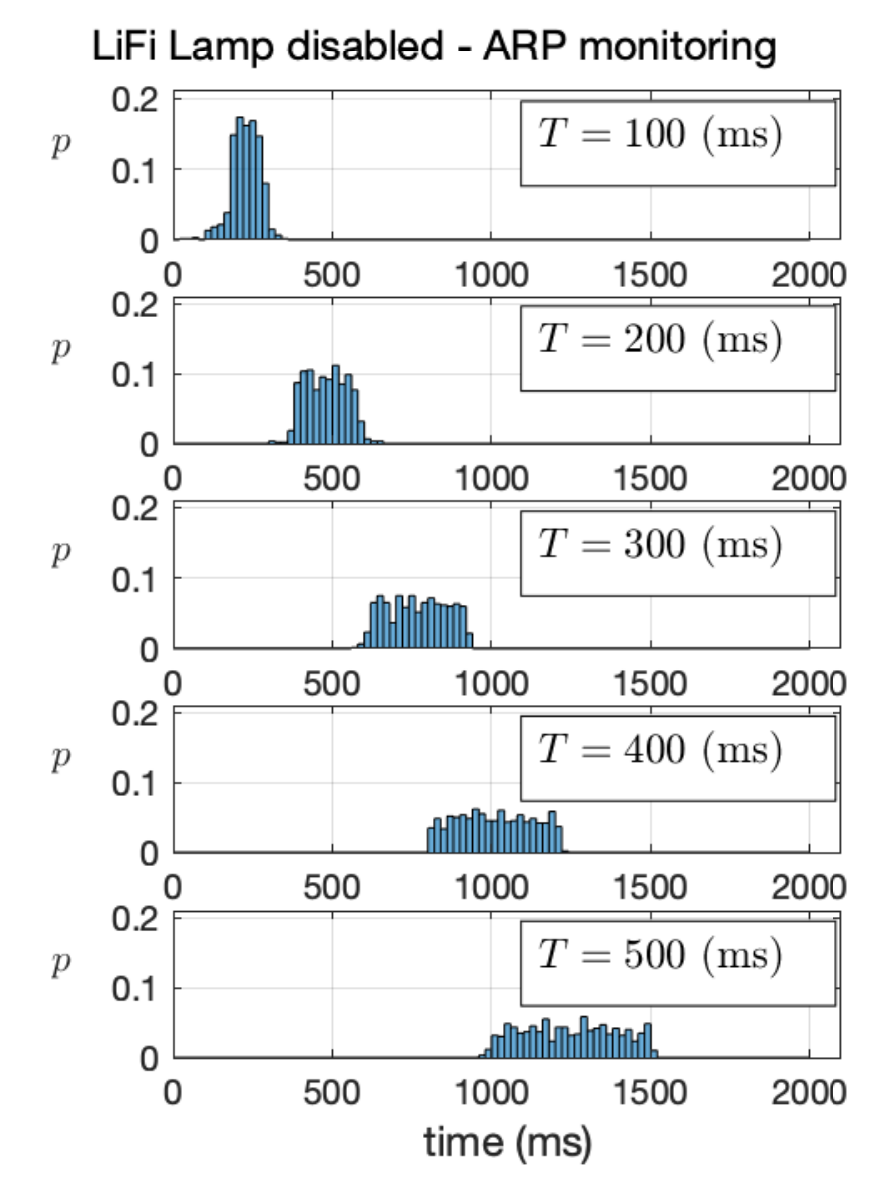}\caption{LiFi}\label{fig:5}\end{subfigure}\begin{subfigure}{0.5\columnwidth}\includegraphics[width=0.95\columnwidth]{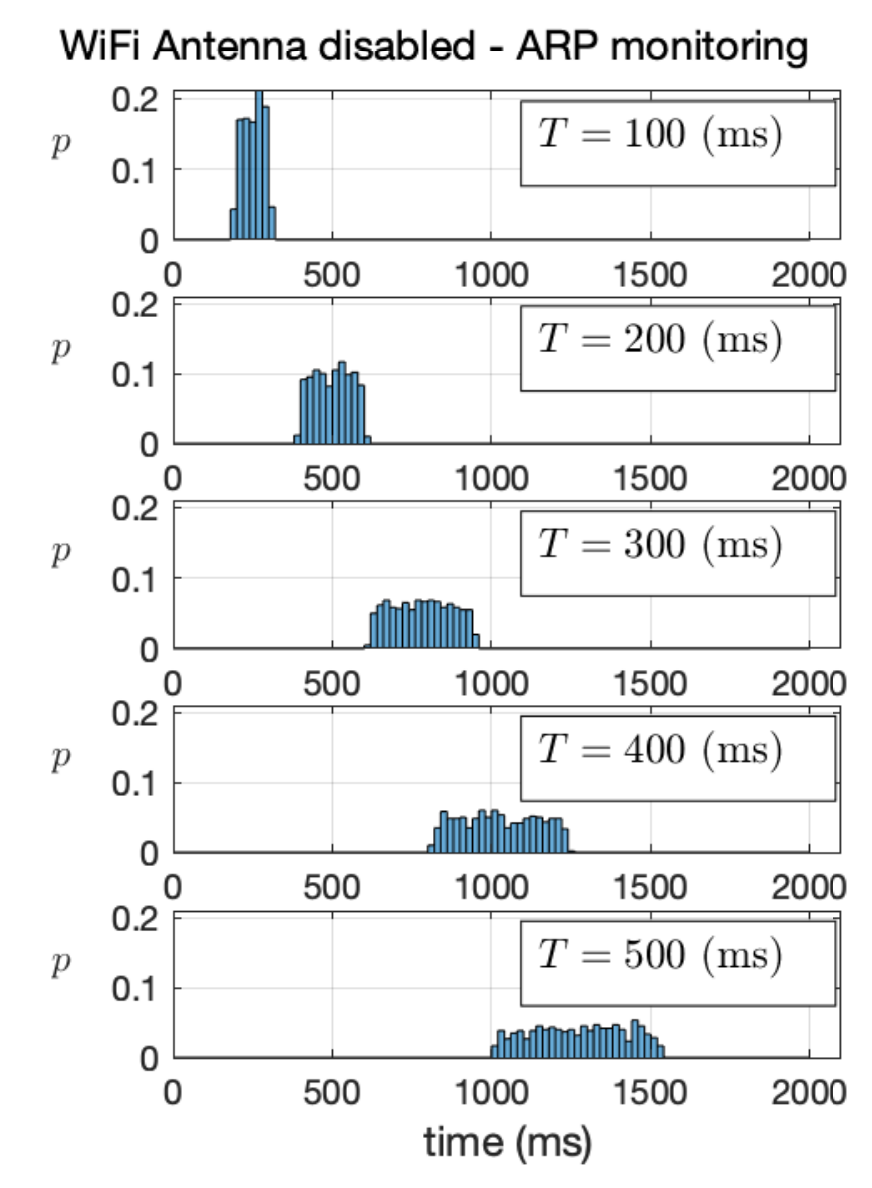}\caption{WiFi}\label{fig:6}\end{subfigure}

\caption{Reaction to a carrier loss with ARP monitoring}
\vspace{-3mm}
\end{figure}

Subfigure~\ref{fig:5} shows that, with the LiFi interface as the
primary one, and ARP monitoring, as the lamp is turned off the system
is able to react, switching to the backup WiFi interface, with a delay
in the range from \textasciitilde 100~ms to 300~ms, when $T$ is
set to 100~ms. The range lower and upper edges linearly increases
to \textasciitilde 1.0~s and \textasciitilde 1.5~s with $T=500$~ms.

Subfigure~\ref{fig:6} shows the results obtained when the system
starts with the WiFi interface and is triggered by a WiFi signal loss
to switch to LiFi. the ARP mechanism with $T=100$~ms is able to
react almost always with a delay between 200~ms and 350~ms. With
$T=500$~ms, the range extends from 1~s to 1.5~s. 

\subsubsection{Switching delay after a signal loss - MII monitoring\label{subsec:Switching-delay-after-signal-loss-MII}}

\begin{figure}[t]
\begin{subfigure}{0.5\columnwidth}\includegraphics[width=0.95\columnwidth]{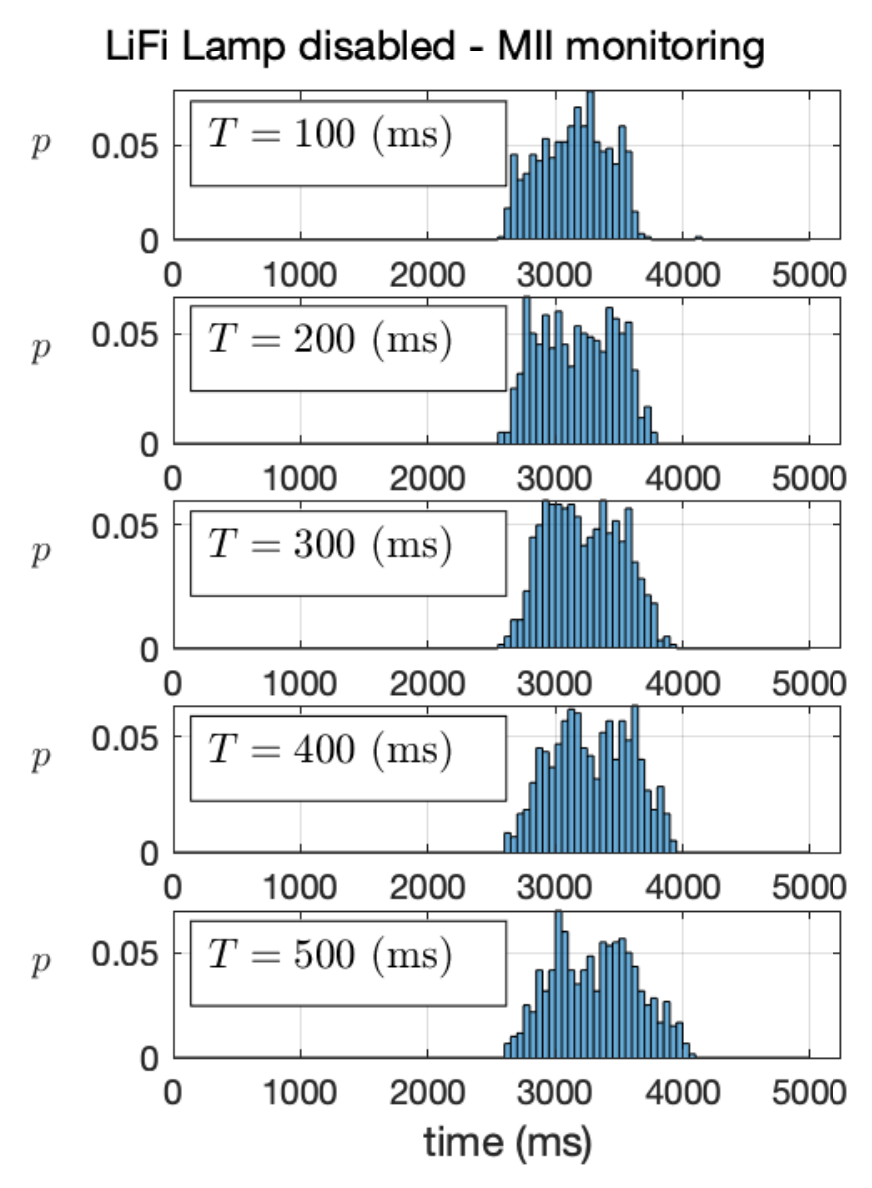}\caption{LiFi}\label{fig:7}\end{subfigure}\begin{subfigure}{0.5\columnwidth}\includegraphics[width=0.95\columnwidth]{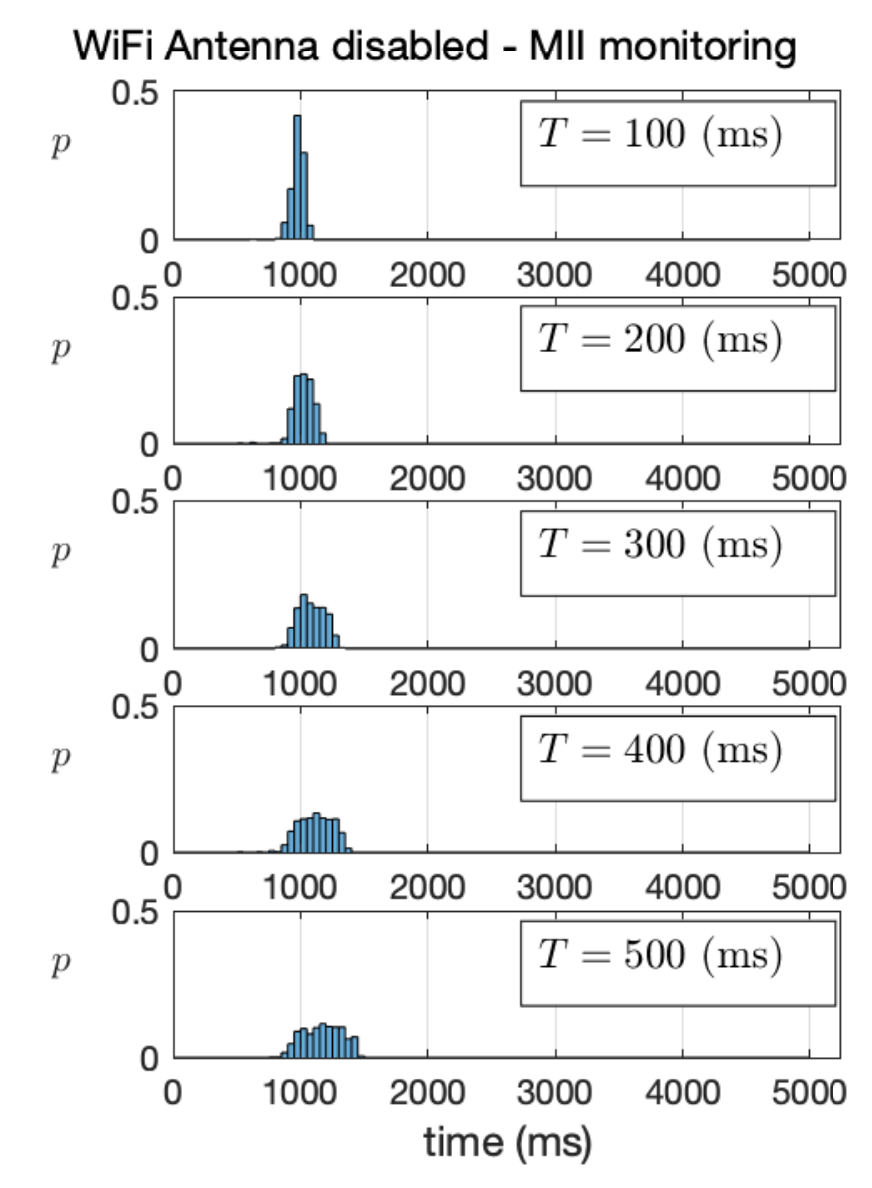}\caption{WiFi}\label{fig:8}\end{subfigure}

\caption{Reaction to a carrier loss with MII monitoring}
\vspace{-6mm}
\end{figure}

In Subfigure~\ref{fig:7} we show the results obtained with MII monitoring
as the LED lamp. in the configuration with LiFi as the primary interface,
is turned off. The event triggers a switch which is completed after
a delay in a range between 2.6~s and \textasciitilde 4~s, with
the upper edge slightly increasing with $T$.

In Subfigure \ref{fig:8} we can see that, with WiFi as the running
interface, turning the power down on the antenna causes an interface
switch to LiFi after a shorter amount of time, ranging between around
\textasciitilde 0.9~s and 1.2~s with $T=100\text{ms}$, and with
the upper edge extending to 1.4~s with $T=500\text{ms}$.

In both subfigures \ref{fig:7} and \ref{fig:8}, the effect of increasing
$T$ is not relevant, almost negligible in figure~\ref{fig:7}. This
behavior tells us that the delay is dominated by the time it takes
to the physical interface to declare the signal as absent. Because
the optical and infrared signals are more subject to environmental
factors with respect to WiFi ones, it is likely that manufacturers
allow for a more conservative amount of time with a poor signal, before
claiming that connectivity is lost.

\subsection{Effect of an intentional interface switch on the traffic flow\vspace{-1mm}}

In the last set of experiments we focused on the effect on a traffic flow,
measured in terms of Packet Loss Percentage (PLR), of switching between
the two interfaces. More precisely, we have checked what happens as
the consequence of an intentional switch, prompted by the operating
systems of the wireless device. In practical scenarios, intentional
interface switches may be the result of the application of an interface
selection policy different from the active-backup policy. For instance,
it may be the result of the decision of some network management agent
operating at upper layers, on the basis or medium or long term network
performance monitoring. Therefore, it is worth quantifying the loss
caused on traffic when this type of decision is taken. For the related
experiments, we set MII monitoring as the method to track the link
status.

To obtain each of the experimental results presented below, 200 replicas
were performed. Each replica consisted in a sending a UDP traffic
stream at 10Mbps in either the downlink or uplink direction to/from
the considered wireless device, for a duration of 40 seconds, starting
the flow with either the LiFi or WiFi as the primary interface. At
the middle of the interval, a primary interface switch command is
sent to the device. Therefore, we have 4 combinations of direction
and primary interface settings. The corresponding results are showed
in Figures 5 and 6. Each subfigure shows the average (solid line)
and corresponding 95\% confidence interval of the PLR in each of the
40 seconds traffic flow in either of the four combinations: Downlink
starting with LiFi as primary, i.e., traffic flows through the LiFi
AP at the beginning and the WiFi AP after switching the primary interface
to WiFi (Subfigure 5a); Uplink starting with LiFi as primary, i.e.,
uplink traffic flows through the LiFi AP at the beginning and the
WiFi AP after switching the primary interface to WiFi (Subfigure 5b);
Downlink starting with WiFi as primary, i.e., traffic flows through
the WiFi AP at the beginning and the LiFi AP after switching the primary
interface to LiFi (Subfigure 6a); and Uplink starting with LiFi as
primary, i.e., traffic flows through the LiFi AP at the beginning
and the WiFi AP after switching the primary interface to WiFi (Subfigure
6b).

In general it can be seen that, when switching occurs from LiFi to
WiFi, packet losses around 0.6\% in the downlink (Subfigure~5a),
and 0.5\% in the uplink (Subfigure~5b), of the traffic flowing during
one second are experienced, which is a already remarkable result.
Switching in the reverse direction, i.e., from WiFi to LiFi, results
in a 0.6\% loss in the downlink (Subfigure~6a), and, notably, 0.15\%
in the uplink (Subfigure~6b).\vspace{-3mm}

\subsection{Discussion\vspace{-1mm}}

Considering the overall picture of the results presented in this paper,
we can conclude that, for a single device, MII monitoring offers superior
performance over ARP monitoring as a consequence of an internal interface
unavailability event. On the other end, in the event of a connectivity
loss, which, in practical scenarios, is supposed to be much more frequent,
ARP monitoring is superior, as it is able to react with an interface
switch in a time below 1 second. Clearly, there is a price to pay
in terms of traffic overhead, as ARP monitoring requires to periodically
sand messages through the network.

Considering specific traffic types, the results obtained seem to suggest
that the considered type of virtual interface, with wireless transmission
technology bonding performed at the dat link layer, is able to support
any time of traffic which does not require stringent (below 1 second)
latency requirements. Web navigation traffic, including streaming
videos (provided that a suitable buffering strategy is in operation)
would be delivered with a satisfactory QoS. On the other hand, video
conferencing traffic, with real-time interactivity, would not be supported
with a sufficient QoS, as a physical connection loss which causes
an interface change would not be recovered for at least 2 seconds.
The considered technology, however, in our view, is promising, if
we consider that the results were obtained without modifying a bonding
driver developed for wired (Ethernet) physical interfaces. The type
of results presented in this work can be the basis to develop proactive
strategies for the physical interface management under a virtual interface,
to pursue a QoS adequate to support all type of traffic. The tradeoff
between traffic overhead and reaction time needs to be investigated
in a multi-user scenario, an objective which we will pursue in our
future work.\texttt{\vspace{-2mm}
}

\begin{figure}[t]
\begin{subfigure}{0.5\columnwidth}\includegraphics[width=0.9\columnwidth]{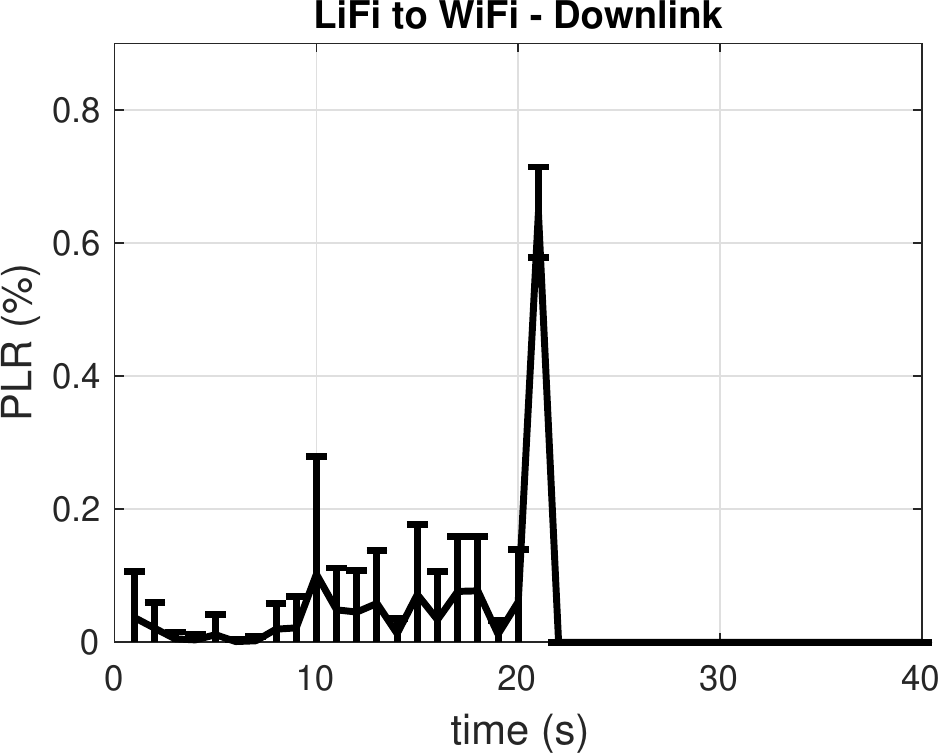}\caption{Downlink}\label{fig:traffic1}\end{subfigure}\begin{subfigure}{0.5\columnwidth}\includegraphics[width=0.9\columnwidth]{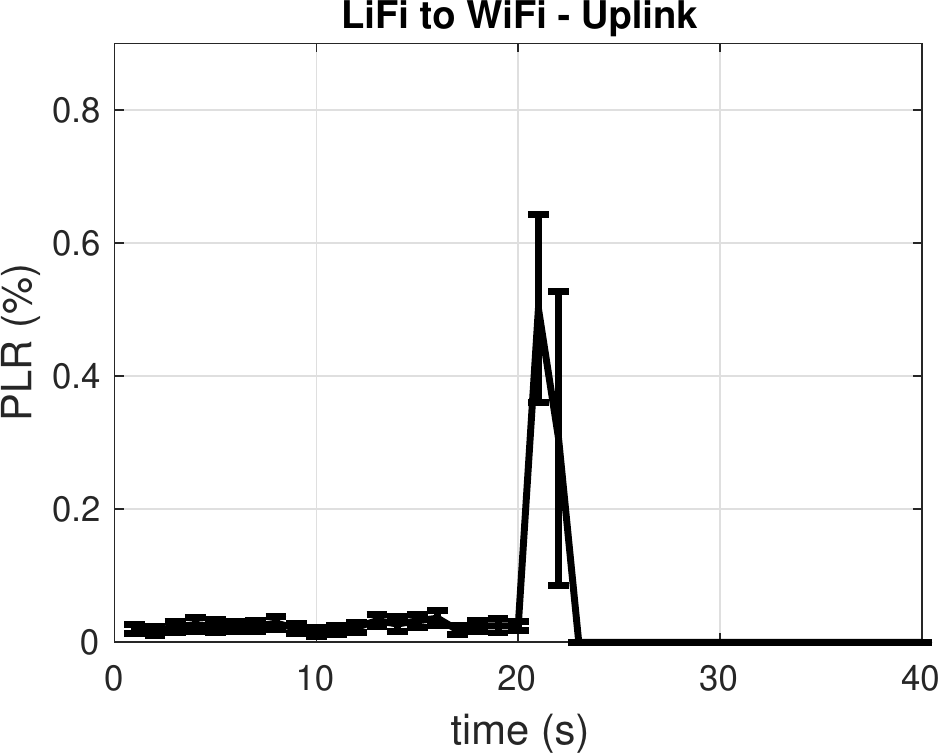}\caption{Uplink}\label{fig:traffic2}\end{subfigure}

\caption{Traffic loss percentage in the transition from LiFi to WiFi}
\texttt{\vspace{-6mm}
}
\end{figure}

\begin{figure}[t]
\begin{subfigure}{0.5\columnwidth}\includegraphics[width=0.9\columnwidth]{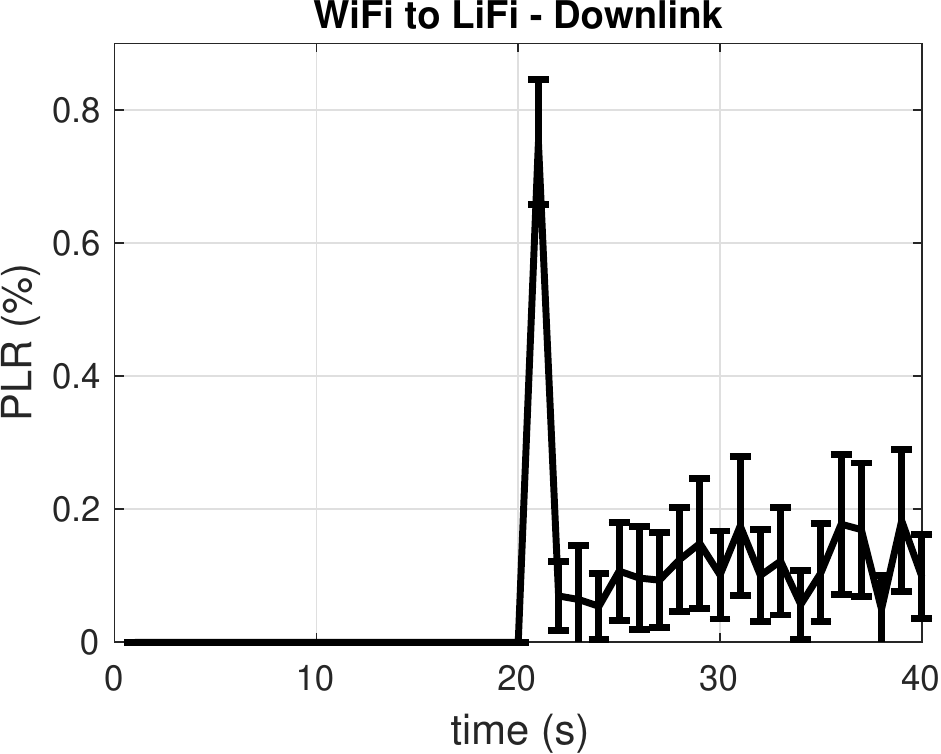}\caption{Downlink}\label{fig:traffic3}\end{subfigure}\begin{subfigure}{0.5\columnwidth}\includegraphics[width=0.9\columnwidth]{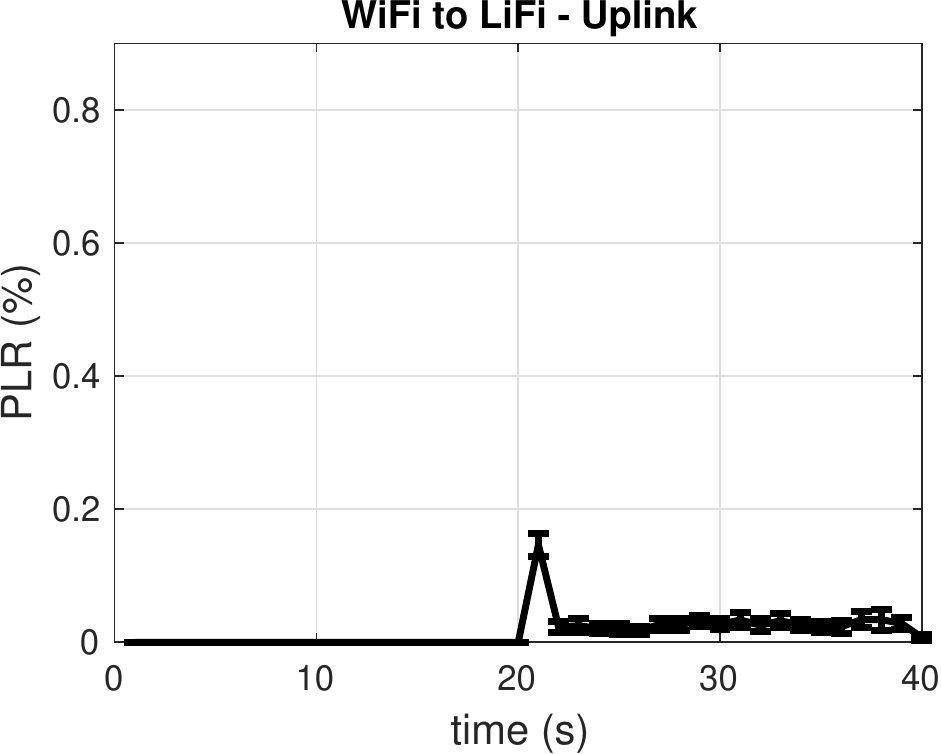}\caption{Uplink}\label{fig:traffic4}\end{subfigure}

\caption{Traffic loss percentage in the transition from WiFi to LiFi}
\texttt{\vspace{-8mm}
}
\end{figure}

\section{Conclusion\label{sec:Conclusion}}

In this work, we have evaluated the performance of a virtual network
interface built on top of LiFi and WiFi interfaces with aggregation
at the data link layer using COTS and the Linux Ethernet Bonding Driver.
Our results show that even using software and hardware tools that
were not originally designed for this purpose, the switching delay
can be kept under two seconds in most of the considered cases when
the switch is caused by an exogenous event (an internal interface
unavailability or a connectivity loss), and the effect on a traffic
flow when an intentional switch is performed is in the order of 1.5\%
lost packets, computed over the packets sent during 1 second. Our
results indicates that the considered type of bonding is able to support
various type of traffic, with the exception of an interactive video
streaming, in which interruptions would be present upon an unintentional
interface switch.\texttt{\vspace{-2mm}
}

\section*{Acknowledgment\texttt{\vspace{-1mm}
}}

This work was funded by the Italian Ministry of University and Research
(MUR) research project AMICO, under grant no. ARS01\_00900.\texttt{\vspace{-2mm}
}

%\bibliographystyle{IEEEtran}
%\bibliography{IEEEabrv,bibliography}

\providecommand{\url}[1]{#1} \csname \endcsname \providecommand{\newblock}{\relax}
\providecommand{\bibinfo}[2]{#2} \providecommand{\BIBentrySTDinterwordspacing}{\spaceskip=0pt\relax}
\providecommand{\BIBentryALTinterwordstretchfactor}{4} \providecommand{\BIBentryALTinterwordspacing}{\spaceskip=\fontdimen2\font plus
\BIBentryALTinterwordstretchfactor\fontdimen3\font minus
  \fontdimen4\font\relax} \providecommand{\BIBforeignlanguage}[2]{{%
\expandafter\ifx\csname l@#1\endcsname\relax
\typeout{** WARNING: IEEEtran.bst: No hyphenation pattern has been}%
\typeout{** loaded for the language `#1'. Using the pattern for}%
\typeout{** the default language instead.}%
\else
\language=\csname l@#1\endcsname
\fi
#2}} \providecommand{\BIBdecl}{\relax} \BIBdecl

\end{document}